\documentclass[aps,twocolumn,groupedaddress,showpacs]{revtex4}
\usepackage{graphicx}%
\usepackage{latexsym}%
\begin{document}
\title{Phase reduction of stochastic limit cycle oscillators}
\author{Kazuyuki Yoshimura and Kenichi Arai} 
\affiliation{
 NTT Communication Science Laboratories, NTT Corporation\\
 2-4, Hikaridai, Seika-cho, Soraku-gun, Kyoto 619-0237, Japan
}
\date{\today}
\begin{abstract}
   We point out that
 the phase reduction of stochastic limit cycle oscillators
 has been done incorrectly in the literature.
 We present a correct phase reduction method
 for oscillators driven by weak external white Gaussian noises.
 Numerical evidence demonstrates that
 the present phase equation properly approximates
 the dynamics of the original full oscillator system.
\end{abstract}
\maketitle
%
%

   Many physical systems can be mathematically modeled
 by limit cycle oscillators.
 It is well known that
 the oscillator systems could exhibit a variety of behaviors. 
 A fundamental theoretical technique for studying the oscillator dynamics
 is the phase reduction method (see e.g. \cite{Kuramoto-1984}).
 This method has been widely and successfully applied to
 coupled oscillators or
 an oscillator subjected to a regular external signal
 such as a periodic one.
 Considerable theoretical progress has been made
 in understanding their dynamics by using this method. 

   Recently,
 the dynamics of oscillators subjected to external stochastic signals
 has also attracted much interest
 in connection with entrainment of independent oscillators
 subjected to a common external noise.
 This common-noise-induced entrainment has been experimentally found
 in several systems as diverse as
 neuronal networks \cite{Mainen-1995},
 ecological systems \cite{Royama-1992},
 and lasers \cite{Yamamoto-2007}. 
 Limit cycle oscillators driven by white Gaussian noise
 have been used
 as simple models for theoretically studying this entrainment
 \cite{Teramae-2004,Goldobin-2005,Nakao-2007,Yoshimura-2007}.
 In these theoretical studies,
 the phase reduction method is applied to the noise-driven oscillators
 to derive a one dimensional equation for the phase variable only.
 Based on this phase equation,
 several reasonable theoretical results have been obtained.
 However, as we will show,
 the phase equation used in the above references is incorrect
 in the sense that in general
 it does not correctly describe
 the dynamics of the original full oscillator system
 even in the weak noise limit.
 
   The phase reduction is a powerful method
 for describing the essential dynamics of oscillators.
 Application field of this method is expected to grow
 also in the case of stochastic oscillators.
 Therefore,
 it is essential to develop a phase reduction method
 for stochastic oscillators.
 In the present paper,
 we consider a general class of limit cycle oscillators,
 which are subjected to white Gaussian noises,
 and develop the phase reduction method for these systems.
 Based on some numerical results,
 it is demonstrated that
 the present phase equation properly approximate
 the dynamics of the original full oscillator system
 while the incorrect version of phase equation fails.
 Finally,
 we make remarks on the results
 concerning the common-noise-induced entrainment,
 which have been obtained in 
 Refs. \cite{Teramae-2004,Goldobin-2005,Nakao-2007,Yoshimura-2007}.
%
%
%

   Let $\mbox{\boldmath $X$}=(x_1,\dots,x_N)\in{\bf R}^N$ be
 a state variable vector and consider the equation
\begin{equation}
 \dot{\mbox{\boldmath $X$}}=
 \mbox{\boldmath $F$}(\mbox{\boldmath $X$})
 +\mbox{\boldmath $G$}(\mbox{\boldmath $X$})\xi(t),
 \label{eqn:limit_cycle_Eq}
\end{equation}
 where
 $\mbox{\boldmath $F$}$ is an unperturbed vector field,
 $\mbox{\boldmath $G$}\in{\bf R}^N$ is a vector function,
 and $\xi(t)$ is the white Gaussian noise
 such that
 $\langle \xi(t) \rangle = 0$ and
 $\langle \xi(t)\xi(s)\rangle = 2D\,\delta(t-s)$,
 where $\langle\cdots\rangle$ denotes
 averaging over the realizations of $\xi$
 and $\delta$ is Dirac's delta function.
 We call the constant $D>0$ the noise intensity.
 The noise-free unperturbed system
 $\dot{\mbox{\boldmath $X$}}
 =\mbox{\boldmath $F$}(\mbox{\boldmath $X$})$
 is assumed to have a limit cycle with a frequency $\omega$.
 We employ the Stratonovich interpretation for
 the stochastic differential equation (\ref{eqn:limit_cycle_Eq}).
 This interpretation allows us to
 use the conventional variable transformations
 in differential equations. 
 A more general form of the noise term has been assumed
 in \cite{Teramae-2004,Goldobin-2005,Nakao-2007}.
 However,
 we assume the form of Eq. (\ref{eqn:limit_cycle_Eq}) for simplicity.
 An extension of the present derivation to a more general case
 is straightforward. 
 
    Consider the unperturbed system
 $\dot{\mbox{\boldmath $X$}}=\mbox{\boldmath $F$}(\mbox{\boldmath $X$})$ 
 and let $\mbox{\boldmath $X$}_0$ be its limit cycle solution.
 A phase coordinate $\phi$ can be defined
 in a neighbourhood $U$ of the limit cycle $\mbox{\boldmath $X$}_0$
 in phase space.
 According to a conventional definition,
 we define the phase variable $\phi$ so that
 $\left(\mbox{grad}_{\mbox{\tiny \boldmath $X$}}\phi\right)
 \cdot\mbox{\boldmath $F$}(\mbox{\boldmath $X$})=\omega$
 may hold for any points in $U$.
 We can define the other $N-1$ coordinates
 $\mbox{\boldmath $r$}=(r_1,\dots,r_{N-1})$ such that
 $\det\partial(\phi,\mbox{\boldmath $r$})/\partial\mbox{\boldmath $X$}\ne 0$
 in $U$.
 We assume that
 $\mbox{\boldmath $r$}=\mbox{\boldmath $a$}$
 on the limit cycle, where
 $\mbox{\boldmath $a$}=(a_1,\dots,a_{N-1})$ is a constant vector.
 If we perform the transformation
 $ (x_1,\dots,x_N)\mapsto(\phi,r_1,\dots,r_{N-1})$
 in Eq. (\ref{eqn:limit_cycle_Eq}),
 we have the equation of the form
\begin{eqnarray}
 \dot{\phi} &=& \omega+h(\phi,\mbox{\boldmath $r$})\xi(t),
 \label{eqn:limit_cycle_Eq_phi}
 \\
 \dot{r}_i &=& f_i(\phi,\mbox{\boldmath $r$})+g_i(\phi,\mbox{\boldmath $r$})\xi(t),
 \label{eqn:limit_cycle_Eq_r}
\end{eqnarray}
 where $i=1,\dots,N-1$.
 The functions $h$, $f_i$, and $g_i$ are defined as follows:
 \begin{eqnarray}
 h(\phi,\mbox{\boldmath $r$}) &=&
 \left(\mbox{grad}_{\mbox{\tiny \boldmath $X$}}\phi\right)
 \cdot\mbox{\boldmath $G$}(\mbox{\boldmath $X$}(\phi,\mbox{\boldmath $r$})),
\\
 f_i(\phi,\mbox{\boldmath $r$}) &=&
 \left(\mbox{grad}_{\mbox{\tiny \boldmath $X$}}r_i\right)
 \cdot\mbox{\boldmath $F$}(\mbox{\boldmath $X$}(\phi,\mbox{\boldmath $r$})),
 \\
 g_i(\phi,\mbox{\boldmath $r$}) &=&
 \left(\mbox{grad}_{\mbox{\tiny \boldmath $X$}}r_i\right)
 \cdot\mbox{\boldmath $G$}(\mbox{\boldmath $X$}(\phi,\mbox{\boldmath $r$})),
 \end{eqnarray}
 where the gradients are evaluated
 at the point $\mbox{\boldmath $X$}(\phi,\mbox{\boldmath $r$})$.
 These functions are periodic with respect to $\phi$: i.e.,
 $h(\phi+2\pi,\mbox{\boldmath $r$})= h(\phi,\mbox{\boldmath $r$})$,
 $f_i(\phi+2\pi,\mbox{\boldmath $r$})= f_i(\phi,\mbox{\boldmath $r$})$,
 and
 $g_i(\phi+2\pi,\mbox{\boldmath $r$})= g_i(\phi,\mbox{\boldmath $r$})$.

   Equations (\ref{eqn:limit_cycle_Eq_phi}) and (\ref{eqn:limit_cycle_Eq_r})
 are Stratonovich stochastic differential equations.
 They can be converted into 
 equivalent Ito stochastic differential equations.
 The $\phi$ component of this Ito type equation
 is obtained as follows:
\begin{eqnarray}
 \dot{\phi} &=&
 \omega+D\biggl[\,
 \frac{\partial h(\phi,\mbox{\boldmath $r$})}{\partial\phi}
 h(\phi,\mbox{\boldmath $r$})
 +\sum_{i=1}^{N-1}\frac{\partial h(\phi,\mbox{\boldmath $r$})}{\partial r_i}
 g_i(\phi,\mbox{\boldmath $r$})
 \,\biggr]
\nonumber
\\
 &&+~ h(\phi,\mbox{\boldmath $r$})\xi(t).
 \label{eqn:Eq_phi_Ito}
\end{eqnarray}
 In the case of weak noise $0<D/\omega\ll 1$,
 the deviation of $\mbox{\boldmath $r$}$ from $\mbox{\boldmath $a$}$
 is expected to be small.
 Thus,
 we can use the approximation 
 $\mbox{\boldmath $r$}=\mbox{\boldmath $a$}$
 in Eq. (\ref{eqn:Eq_phi_Ito}).
 Using this approximation, we arrive at
\begin{equation}
 \dot{\phi} =
 \omega+D\left[\,Z'(\phi)Z(\phi)+W(\phi)\,\right]
 +Z(\phi)\xi(t),
 \label{eqn:phase_Eq}
\end{equation}
 where $Z(\phi)$ and $W(\phi)$ are given by
\begin{eqnarray}
 Z(\phi) &=& h(\phi,\mbox{\boldmath $a$}),
 \label{eqn:def_Z}
\\
 W(\phi) &=&
 \sum_{i=1}^{N-1}\frac{\partial h(\phi,\mbox{\boldmath $a$})}{\partial r_i}
 g_i(\phi,\mbox{\boldmath $a$}),
 \label{eqn:def_W}
\end{eqnarray}
 respectively.
 Since $h$ and $g_i$ are periodic functions,
 $Z(\phi)$ and $W(\phi)$ are also periodic:
 i.e., $Z(\phi+2\pi)=Z(\phi)$ and $W(\phi+2\pi)=W(\phi)$.
 We may conclude that
 the reduced phase equation for the noise-driven oscillator
 (\ref{eqn:limit_cycle_Eq}) is given by
 Eq. (\ref{eqn:phase_Eq}).
 The oscillator dynamics is often studied
 by assuming a phase model instead of
 multidimensional differential equations.
 We emphasize that
 an equation of the form (\ref{eqn:phase_Eq})
 has to be assumed in studying
 the dynamics of oscillators with white Gaussian noises.

   In the previous studies
 \cite{Teramae-2004,Goldobin-2005,Nakao-2007,Yoshimura-2007},
 the authors assumed the Ito type reduced phase equation
 of the form
\begin{equation}
 \dot{\phi}=\omega+DZ(\phi)Z'(\phi)+Z(\phi)\xi(t).
 \label{eqn:incorrect_phase_Eq}
\end{equation}
 Comparison of the present phase equation (\ref{eqn:phase_Eq})
 and Eq. (\ref{eqn:incorrect_phase_Eq})
 clearly shows that
 the term $DW(\phi)$ is dropped
 in the previously used equation (\ref{eqn:incorrect_phase_Eq}).
 This term is $O(D)$ and is of the same order as $DZ(\phi)Z'(\phi)$.
 Thus,
 in general,
 equation (\ref{eqn:incorrect_phase_Eq})
 does not correctly describe
 the essential dynamics of the oscillator
 even in the lowest order approximation
 as will be demonstrated.
 In the exceptional case $W\simeq 0$,
 it gives reasonable results.
%
%
%

   The reduced phase equation (\ref{eqn:phase_Eq}) is
 useful to calculate statistical quantities,
 which characterize the dynamics of
 oscillators subjected to white Gaussian noises.
 Using Eq. (\ref{eqn:phase_Eq}),
 we derive analytical expressions for
 two fundamental statistical quantities,
 which are
 the steady probability distribution of the phase variable
 and the mean frequency $\Omega$.
 We will compare these quantities
 obtained by using a two dimensional oscillator model
 with
 those obtained by using its reduced phase model.

   Let $P(\phi,t)$ be
 the time-dependent probability distribution function
 for the phase $\phi$.
 The stochastic differential equation (\ref{eqn:phase_Eq})
 is equivalent to the Fokker-Planck equation
\begin{eqnarray}
 \frac{\partial P}{\partial t} &=&
 -\frac{\partial }{\partial \phi}
 \bigl[\{\omega+DZ(\phi)Z'(\phi)+DW(\phi)\}P\bigr]
\nonumber
\\
 &&+D\frac{\partial^2}{\partial\phi^2}\bigl[Z(\phi)^2P\bigr].
 \label{eqn:FP_Eq_phi}
 \end{eqnarray}
 We consider Eq. (\ref{eqn:FP_Eq_phi}) over the interval $\phi\in[0,2\pi]$
 and assume the periodic boundary condition $P(0,t)=P(2\pi,t)$.
 The steady solution $P(\phi)$ is obtained
 by assuming $\partial P/\partial t=0$ in Eq. (\ref{eqn:FP_Eq_phi}).
 If we construct
 an asymptotic solution for $P(\phi)$
 in the power of $\varepsilon\equiv D/\omega$,
 then up to the first order we can obtain
\begin{equation}
 P(\phi) =
 \frac{1}{2\pi}
 + \frac{\varepsilon}{2\pi}
 \left[\,Z(\phi)Z'(\phi)-W(\phi)+\overline{W}\,\right]+O(\varepsilon^2),
 \label{eqn:P_1st_order}
\end{equation}
 where $\overline{W}$ is defined by
 $\overline{W}=(2\pi)^{-1}\int_{0}^{2\pi}W(\phi)d\phi$.

   The mean frequency $\Omega$ of the oscillator
 is defined by
\begin{equation}
 \Omega =
 \lim_{T\rightarrow\infty}\frac{1}{T}\int_{0}^{T}\dot{\phi}(t) dt.
 \label{eqn:def_mean-frequency}
\end{equation}
 This can be calculated
 by replacing the time average (\ref{eqn:def_mean-frequency})
 with the ensemble average:
 i.e., $\Omega=\langle\dot{\phi}\rangle$.
 In the Ito equation, unlike in Stratonovich formulation,
 the correlation between $\phi$ and $\xi$ vanishes.
 If we take the ensemble average of Eq. (\ref{eqn:phase_Eq}),
 then we have
\begin{equation}
 \langle\dot{\phi}\rangle =
 \omega+D\langle\,Z(\phi)Z'(\phi)+W(\phi)\,\rangle,
 \label{eqn:average_phase_Eq_Ito}
\end{equation}
 where we used the fact $\langle Z(\phi)\xi(t) \rangle
 = \langle Z(\phi) \rangle\langle \xi(t) \rangle =0$.
 For an arbitrary function $A(\phi)$,
 the ensemble average can be calculated by
 using the steady probability distribution $P(\phi)$:
 i.e., $\langle\, A \,\rangle =\int_{0}^{2\pi}A(\phi)P(\phi)d\phi$.
 If we use Eq. (\ref{eqn:P_1st_order}),
 we can obtain $\Omega$ up to the second order in $\varepsilon$ as follows:
\begin{equation}
 \frac{\Omega}{\omega} =
 1 + \varepsilon\,\overline{W}
 +\varepsilon^2\left[\,
 \overline{(ZZ')^2}-\overline{W^2}+\overline{W}^{\,2}
 \,\right]
 +O(\varepsilon^3),
 \label{eqn:omega_2nd_order}
\end{equation}
 where
 $\overline{(ZZ')^2}=
 (2\pi)^{-1}\int_{0}^{2\pi}\{Z(\phi)Z'(\phi)\}^2d\phi$
 and $\overline{W^2}=
 (2\pi)^{-1}\int_{0}^{2\pi}\{W(\phi)\}^2d\phi$.
 Since the white Gaussian noise has no characteristic frequency,
 intuitively, one might expect that
 the noise does not cause any change in the oscillator frequency.
 However, this is not the case.
 Equation (\ref{eqn:omega_2nd_order}) clearly shows that
 an external white Gaussian noise changes
 the mean frequency $\Omega$ in a general class of oscillators.
 It depends on the sign of $\overline{W}$
 whether $\Omega$ increases or decreases
 as the noise intensity increases.
 
   Equations (\ref{eqn:P_1st_order}) and (\ref{eqn:omega_2nd_order})
 show that
 the term $W(\phi)$ in Eq. (\ref{eqn:phase_Eq}) significantly affects
 both the steady probability distribution $P(\phi)$ and
 the mean frequency $\Omega$
 in the first order of $\varepsilon$.
 In particular, as shown by Eq. (\ref{eqn:omega_2nd_order}),
 the first order frequency shift is determined only from $W(\phi)$.
 Therefore,
 it is crucially important to include the term $W(\phi)$
 into the reduced phase equation as in Eq. (\ref{eqn:phase_Eq}) .
 It is clear that
 the previously used phase equation (\ref{eqn:incorrect_phase_Eq})
 cannot give proper approximations for $P(\phi)$ and $\Omega$.
%
%
%

   In order to validate the above phase reduction method, 
 we carried out numerical calculations for
 an example of noise-driven oscillator.
 We compare  $P(\phi)$ and $\Omega$
 between the theoretical and numerical results.
 We consider the Stuart-Landau (SL) oscillator
\begin{eqnarray}
 \dot{x} &=& x-c_{0}y-(x^2+y^2)(x-c_{2}y)+G_x\xi(t),
 \label{eqn:Stuart-Landau_osc_x}
\\
 \dot{y} &=& c_{0}x+y-(x^2+y^2)(c_{2}x+y)+G_y\xi(t),
 \label{eqn:Stuart-Landau_osc_y}
\end{eqnarray}
 where $c_0$ and $c_2$ are constants,
 $\mbox{\boldmath $G$}=(G_x,G_y)$ is a vector function of $(x,y)$,
 and $\xi$ is the white Gaussian noise
 with the properties
 $\langle \xi(t) \rangle = 0$ and
 $\langle \xi(t)\xi(s)\rangle = 2D\,\delta(t-s)$.
 The SL oscillator has the limit cycle solution
 $\mbox{\boldmath $X$}_0=(\cos\omega t,\sin\omega t)$,
 where the natural frequency $\omega$ is given by $\omega=c_0-c_2$. 
 If we define the coordinates $(\phi,r)$ by the transformation
\begin{equation}
 x=r\cos(\phi+c_2\ln r),\quad
 y=r\sin(\phi+c_2\ln r),
 \label{eqn:def_phi-r_SL}
\end{equation}
 then $\phi$ gives the phase variable
 and the limit cycle is represented by $r=1$.
 
    The functions $Z(\phi)$ and $W(\phi)$ can be obtained
 from Eq. (\ref{eqn:def_phi-r_SL}).
 As examples,
 we consider the following two types of $\mbox{\boldmath $G$}$:
 $\mbox{\boldmath $G$}_1=(1,0)$
 and $\mbox{\boldmath $G$}_2=(x,0)$.
 For the first example,
 $Z(\phi)$ and $W(\phi)$  are given by
 $Z(\phi) = -\left(\sin\phi+c_{2}\cos\phi\right)$
 and $W(\phi) = \{(1+c_{2}^2)/2\}\sin2\phi$.
 For the second example, they are given by
 $Z(\phi) = -\cos\phi\left( \sin\phi+c_{2}\cos\phi \right)$
 and $W(\phi) = c_{2}\cos^2\phi\left( -\cos2\phi+c_{2}\sin2\phi \right)$.
 Approximations for $P(\phi)$ and $\Omega$ can be obtained
 by substituting these expressions for $Z(\phi)$ and $W(\phi)$
 into Eqs. (\ref{eqn:P_1st_order}) and (\ref{eqn:omega_2nd_order}).
 As for $\Omega$,
 we can obtain $\Omega/\omega=1+O(\varepsilon^3)$
 for the first example $\mbox{\boldmath $G$}_1$
 and $\Omega/\omega
 =1-(c_2/4)\varepsilon+\{(1+c_{2}^2)/32\}\varepsilon^2+O(\varepsilon^3)$
 for the second example $\mbox{\boldmath $G$}_2$.
 The former indicates that
 $\Omega$ is independent of $c_2$ and
 constant up to the second order in the first example.
 In contrast,
 the latter indicates that
 $\Omega$ can either increase or decrease in the first order
 depending on $c_{2}$ in the second example.
 
\begin{figure}[t]
 \includegraphics[width=85mm,height=65mm]{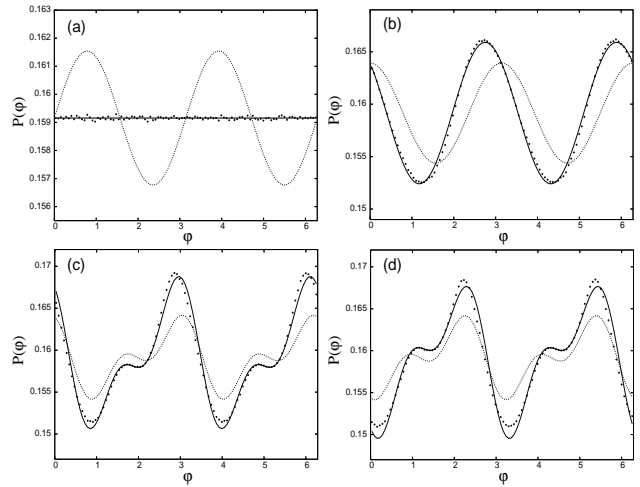}
 \caption{Steady probability distribution $P(\phi)$ of phase
 for noise-driven SL oscillator.
 Numerical result ($\bullet$),
 analytical result Eq. (\ref{eqn:P_1st_order}) (solid line),
 and that obtained from Eq. (\ref{eqn:incorrect_phase_Eq}) (dashed line)
 are shown for $\varepsilon=0.03$.
 (a) $\mbox{\boldmath $G$}_1$ and $(c_0,c_2)=(1,0)$;
 (b) $\mbox{\boldmath $G$}_1$ and $(c_0,c_2)=(2,1)$;
 (c) $\mbox{\boldmath $G$}_2$ and $(c_0,c_2)=(2,1)$;
 (d) $\mbox{\boldmath $G$}_2$ and $(c_0,c_2)=(0,-1)$.
 }
 \label{fig:steady_distribution}
\end{figure}

   In Figs. \ref{fig:steady_distribution}(a)-(d),
 numerical and theoretical results for $P(\phi)$
 are compared:
 filled circle and solid line represent
 $P(\phi)$ obtained by numerically solving
 Eqs. (\ref{eqn:Stuart-Landau_osc_x}) and (\ref{eqn:Stuart-Landau_osc_y})
 and that given by Eq. (\ref{eqn:P_1st_order}), respectively.
 Theoretical predictions made by Eq. (\ref{eqn:incorrect_phase_Eq}),
 which are obtained just by setting $W=0$ in Eq. (\ref{eqn:P_1st_order}),
 are also shown by dashed line.
 Figures \ref{fig:steady_distribution}(a) and (b) are for
 the case of $\mbox{\boldmath $G$}_1$
 while figures \ref{fig:steady_distribution}(c) and (d) are for
 the case of $\mbox{\boldmath $G$}_2$.
 It is clear that
 the present phase model (\ref{eqn:phase_Eq}) gives
 precise approximations in all the cases.
 The agreements are excellent.
 In contrast,
 the incorrect version of phase equation (\ref{eqn:incorrect_phase_Eq})
 does not give proper approximations at all
 in spite of the weak noise intensity. 

\begin{figure}[t]
 \includegraphics[width=85mm,height=33mm]{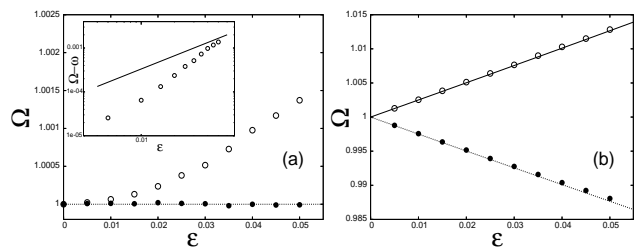}
 \caption{Mean frequency $\Omega$ vs. $\varepsilon$
 for (a) $\mbox{\boldmath $G$}_1$ and (b) $\mbox{\boldmath $G$}_2$.
 Numerical result (symbol) and
 analytical result Eq. (\ref{eqn:omega_2nd_order}) (line) are shown.
 (a) $(c_0,c_2)=(1,0)$ ($\bullet$, dashed line) and
      $(c_0,c_2)=(2,1)$ ($\circ$, dashed line);
 (b) $(c_0,c_2)=(2,1)$ ($\bullet$, dashed line) and
      $(c_0,c_2)=(0,-1)$ ($\circ$, solid line).
 The inset in (a) is
 logarithmic plot of $\Omega-\omega$ vs. $\varepsilon$,
 where reference line for
 the scaling law $\varepsilon^1$ is also shown.
 }
 \label{fig:mean-frequency}
\end{figure}

   Figures \ref{fig:mean-frequency}(a) and (b) show
 the mean frequency $\Omega$
 plotted as a function of $\varepsilon=D/\omega$ for
 the cases of $\mbox{\boldmath $G$}_1$ and $\mbox{\boldmath $G$}_2$,
 respectively.
 In all the numerical calculations,
 the natural frequency is set as $\omega=1$.
 The numerical results obtained by solving
 Eqs. (\ref{eqn:Stuart-Landau_osc_x}) and (\ref{eqn:Stuart-Landau_osc_y})
 are shown by filled or open circle.
 The theoretical estimations given by Eq. (\ref{eqn:omega_2nd_order})
 are also shown by solid or dashed line.
 The theoretical estimation is $\Omega/\omega=1+O(\varepsilon^3)$
 for $\mbox{\boldmath $G$}_1$,
 which is independent of $\varepsilon$ and
 constant up to the second order in $\varepsilon$.
 In Fig.  \ref{fig:mean-frequency}(a),
 the numerically obtained $\Omega$ is almost constant
 for $(c_0,c_2)=(1,0)$.
 This coincides with the above theoretical estimation.
 In the case of $(c_0,c_2)=(2,1)$, 
 there is a deviation between the numerical and theoretical results:
 the numerical result shows an increase with increasing $\varepsilon$.
 However,
 this increase in not linear with respect to $\varepsilon$
 but a higher order one as shown in the inset.
 In this sense,
 an agreement between the numerical and theoretical results
 is confirmed up to the first order.
 In the case of $\mbox{\boldmath $G$}_2$,
 the theoretical estimation is given by
 $\Omega/\omega
 =1-(c_2/4)\varepsilon+\{(1+c_{2}^2)/32\}\varepsilon^2+O(\varepsilon^3)$,
 which has a non-vanishing term of $O(\varepsilon)$
 except for $c_2=0$.
 In Fig.  \ref{fig:mean-frequency}(b),
 a good agreement between the numerical result and this estimation
 is obtained
 in each of the cases $(c_0,c_2)=(2,1)$ and $(0,-1)$.
 If we use Eq. (\ref{eqn:incorrect_phase_Eq}) instead of Eq. (\ref{eqn:phase_Eq}),
 then we obtain the estimation $\Omega/\omega=1+O(\varepsilon^2)$,
 in which the $O(\varepsilon)$ term vanishes.
 This  estimation apparently disagrees with the numerical results.
 
    Figures. \ref{fig:steady_distribution} and \ref{fig:mean-frequency}
 clearly demonstrate that
 the reduced phase equation (\ref{eqn:phase_Eq})
 precisely approximate the dynamics of stochastic oscillators
 with weak white Gaussian noises.
 In addition,
 it is apparent that
 the previously used equation (\ref{eqn:incorrect_phase_Eq})
 is erroneous.
%
%
%
%

   It is known that
 an entrainment could occur between
 two independent oscillators
 subjected to a common external white Gaussian noise.
 We discuss this entrainment phenomenon,
 paying particular attention to an effect of the term $W(\phi)$.
 Consider the two equations
\begin{equation}
 \dot{\mbox{\boldmath $X$}_i}=
 \mbox{\boldmath $F$}(\mbox{\boldmath $X$}_i)
 +\delta\mbox{\boldmath $F$}_i(\mbox{\boldmath $X$}_i)
 +\mbox{\boldmath $G$}_i(\mbox{\boldmath $X$}_i)\xi(t),
 \quad i=1,2,
 \label{eqn:cnis_limit_cycle_Eq}
\end{equation}
 where
 $\mbox{\boldmath $X$}_i \in{\bf R}^N$,
 $\mbox{\boldmath $F$}$ is an unperturbed vector field,
 $\delta\mbox{\boldmath $F$}_1$ and
 $\delta\mbox{\boldmath $F$}_2$ are small deviations from it,
 $\mbox{\boldmath $G$}_1$ and $\mbox{\boldmath $G$}_2$
 are slightly different vector functions,
 and $\xi(t)$ is the common white Gaussian noise with
 $\langle \xi(t) \rangle = 0$ and
 $\langle \xi(t)\xi(s)\rangle = 2D\,\delta(t-s)$.
 The phase is defined by the unperturbed system
 $\dot{\mbox{\boldmath $X$}}
 =\mbox{\boldmath $F$}(\mbox{\boldmath $X$})$.
 Equation (\ref{eqn:cnis_limit_cycle_Eq})
 can be reduced into the phase equation
\begin{equation}
 \dot{\phi}_i =
 \omega_i+D\left[\,Z_i'(\phi_i)Z_i(\phi_i)+W_i(\phi_i)\,\right]
 +Z_i(\phi_i)\xi(t),
 \label{eqn:cnis_phase_Eq}
\end{equation}
 where $\omega_i$ represents the natural frequency.
 For simplicity,
 we assume $\omega_i$ is a constant. 
 We introduce the average $Z(\phi)=(Z_1(\phi)+Z_2(\phi))/2$
 and the difference $\delta Z(\phi)=(Z_1(\phi)-Z_2(\phi))/2$.
 The difference $\delta Z(\phi)$ is small
 since
 $\mbox{\boldmath $G$}_1\simeq \mbox{\boldmath $G$}_2$
 is assumed.
 
   Let $\theta$ and $\psi$ be defined by
 $\theta=\phi_1-\phi_2$ and $\psi=\phi_1+\phi_2-2\omega t$,
 where $\omega$ is the average natural frequency
 defined by $\omega=(\omega_1+\omega_2)/2$.
 The variable $\theta$ measures the phase difference
 between the two oscillators.
 In the case of weak noise,
 $\theta$ and $\psi$ can be regarded as slow variables
 and thus the averaging approximation can be applied.
 If we perform the time-averaging and
 neglect the terms which are of the order of $D|\delta Z|$,
 then we can obtain the Fokker-Planck equation for
 the probability distribution $Q(t,\theta,\psi)$ as follows:
\begin{eqnarray}
 \frac{\partial Q}{\partial t} \!\!&=&\!\!
 -\left\{\delta\omega+D\left(\overline{W}_1-\overline{W}_2\right)\right\}
  \frac{\partial Q}{\partial \theta}
 \!-D\left(\overline{W}_1+\overline{W}_2\right)\frac{\partial Q}{\partial \psi}~
\nonumber
\\
 &&\!\!+ D\frac{\partial^2}{\partial\theta^2}[u(\theta)Q]
 +D\frac{\partial^2}{\partial\psi^2}[v(\theta)Q],
 \label{eqn:cnis_FP_Eq}
\end{eqnarray}
 where $\delta\omega=\omega_1-\omega_2$.
 The functions $u$ and $v$ are defined as
 $u(\theta)=2\{{\mit\Gamma}(0)-{\mit\Gamma}(\theta)\}$
 and $v(\theta)=2\{{\mit\Gamma}(0)+{\mit\Gamma}(\theta)\}$,
 where
 ${\mit\Gamma}(\theta)=(2\pi)^{-1}\int_{0}^{2\pi}Z(\phi)Z(\phi+\theta)d\phi$.
 Equation (\ref{eqn:cnis_FP_Eq}) has a steady solution $Q(\theta)$,
 which is a function of $\theta$ only.
 The entrainment phenomenon is characterized by $Q(\theta)$.
 The steady solution is determined by the equation
\begin{eqnarray}
 D\frac{d[u(\theta)Q]}{d\theta}
 -\left\{\delta\omega+D\left(\overline{W}_1-\overline{W}_2\right)\right\}Q =C,
 \label{eqn:cnis_steady_FP_Eq}
\end{eqnarray}
 where $C$ is an integration constant.
 In Eq. (\ref{eqn:omega_2nd_order}),
 it has been shown that
 the mean frequency $\Omega$ shifts
 by $D\overline{W}$ in the lowest order.
 Equation (\ref{eqn:cnis_steady_FP_Eq}) indicates that
 this frequency shift effect appears
 as the effective detuning $D(\overline{W}_1-\overline{W}_2)$.
 The profile of $Q(\theta)$ depends on the coefficient of $Q$
 in Eq. (\ref{eqn:cnis_steady_FP_Eq}).
 It has been shown that
 $Q(\theta)$ has peaks at the zero points of $u(\theta)$
 and these peaks become narrower and higher,
 which corresponds to better synchronization quality,
 as the ratio
 $\{\delta\omega+D(\overline{W}_1-\overline{W}_2)\}/D$
 between the coefficient of $Q$ and $D$
 becomes smaller
 \cite{Yoshimura-2007}.
 Therefore,
 the profile of $Q(\theta)$ depends on
 the functional form of $W_i(\phi)$ in Eq. (\ref{eqn:cnis_phase_Eq}).
 It may be concluded that
 the contribution due to $W(\phi)$ in Eq. (\ref{eqn:phase_Eq})
 is not negligible in the common-noise-induced entrainment.
 In addition,
 equation (\ref{eqn:cnis_steady_FP_Eq}) suggests
 that the synchronization quality could be improved
 if $\delta\omega$ and $D(\overline{W}_1-\overline{W}_2)$
 cancel with each other.
 
   The previous works
 \cite{Teramae-2004,Goldobin-2005,Nakao-2007,Yoshimura-2007}
 have assumed the case
 $\mbox{\boldmath $G$}_1=\mbox{\boldmath $G$}_2$.
 In this particular case,
 $W_1=W_2$ holds and thus
 the effective detuning $D(\overline{W}_1-\overline{W}_2)$ vanishes
 in Eq. (\ref{eqn:cnis_steady_FP_Eq}).
 Therefore,
 the same equation for $Q(\theta)$ can be obtained
 even if $W_i(\phi)$ in Eq. (\ref{eqn:cnis_phase_Eq})
 is not taken into account.
 Because of this fact, fortunately,
 an analysis based on
 the incorrect phase equation (\ref{eqn:incorrect_phase_Eq})
 also leads to correct results.
%
%
%

   In conclusion,
 we have developed
 the phase reduction method for
 a general class of limit cycle oscillators
 subjected to white Gaussian noises.
 Applying the present reduced phase equation,
 we derived analytical expressions for
 the steady probability distribution $P(\phi)$ of phase
 and the mean frequency $\Omega$.
 It has been found that
 an external white Gaussian noise gives rise to a frequency shift.
 We showed that
 these analytical estimations of $P(\phi)$ and $\Omega$
 are in good agreement with numerical results
 to demonstrate that
 the present phase equation properly approximates
 the dynamics of the original full oscillator system.
 In addition,
 we pointed out that
 an effect due to the frequency shift emerges
 also in the common-noise-induced entrainment.
%
%
%

   The authors would like to thank 
 the members of NTT Communication Science Laboratories
 for their continual encouragements.
%
%
%
%

%
%
\end{document}